\documentclass[a4paper,twocolumn,11pt,accepted=2024-01-22]{quantumarticle}
\pdfoutput=1
\usepackage[utf8]{inputenc}
\usepackage[english]{babel}
\usepackage[T1]{fontenc}
\usepackage{amsmath}
\usepackage{amssymb}
\usepackage{multirow}
\usepackage{qcircuit}
\usepackage{physics}
\usepackage{caption}
\usepackage{graphicx}
\usepackage{xcolor}
\usepackage{soul}
\usepackage{hyperref}

\usepackage{tikz}
\usepackage{lipsum}

\begin{document}

\title{Digital quantum simulation of non-perturbative dynamics of open systems with orthogonal polynomials}

\author{José D. Guimarães}
\affiliation{Centro de F\'{\i}sica das Universidades do Minho e do Porto, Braga 4710-057, Portugal}
\affiliation{Institute of Theoretical Physics and IQST, Ulm University, Albert-Einstein-Allee 11, Ulm 89081, Germany}
 \affiliation{International Iberian Nanotechnology Laboratory, Av. Mestre Jos\'e Veiga s/n, Braga 4715-330, Portugal}
 \email{jose.diogo-da-costa@uni-ulm.de}
\author{Mikhail I. Vasilevskiy}
\affiliation{International Iberian Nanotechnology Laboratory, Av. Mestre Jos\'e Veiga s/n, Braga 4715-330, Portugal}
\affiliation{Laboratório de Física para Materiais e Tecnologias Emergentes (LaPMET), Universidade do Minho, Braga 4710-057, Portugal}
\affiliation{Departamento de F\'{\i}sica, Universidade do Minho, Braga 4710-057, Portugal}
\author{Lu\'is S. Barbosa}
     \affiliation{International Iberian Nanotechnology Laboratory, Av. Mestre Jos\'e Veiga s/n, Braga 4715-330, Portugal}
    \affiliation{INESC TEC, Departamento de Informática, Universidade do Minho, Braga 4710-057, Portugal}
\maketitle

\begin{abstract}
  Classical non-perturbative simulations of open quantum systems' dynamics face several scalability problems, namely, exponential scaling of the computational effort as a function of either the time length of the simulation or the size of the open system. In this work, we propose the use of the Time Evolving Density operator with Orthogonal Polynomials Algorithm (TEDOPA) on a quantum computer, which we term as Quantum TEDOPA (Q-TEDOPA), to simulate non-perturbative dynamics of open quantum systems linearly coupled to a bosonic environment (continuous phonon bath). By performing a change of basis of the Hamiltonian, the TEDOPA yields a chain of harmonic oscillators with only local nearest-neighbour interactions, making this algorithm suitable for implementation on quantum devices with limited qubit connectivity such as superconducting quantum processors. We analyse in detail the implementation of the TEDOPA on a quantum device and show that exponential scalings of computational resources can potentially be avoided for time-evolution simulations of the systems considered in this work. We applied the proposed method to the simulation of the exciton transport between two light-harvesting molecules in the regime of moderate coupling strength to a non-Markovian harmonic oscillator environment on an IBMQ device. Applications of the Q-TEDOPA span problems which can not be solved by perturbation techniques belonging to different areas, such as the dynamics of quantum biological systems and strongly correlated condensed matter systems.
\end{abstract}
\section{Introduction} \label{sec:outline}
The emergence of decoherence in quantum systems is an important and ubiquitous phenomenon in Nature because any real system is never completely isolated. On the other hand, Hamiltonian dynamics simulation of such a system is by no means trivial, since its environment typically involves an intractable number of degrees of freedom. Approximations are thus usually employed, however, they limit the range of applications of the computational method. Hence, non-perturbative methods have been proposed to be applied to general open quantum systems, in particular, those where perturbative techniques fail. One of such techniques is the Hierarchical Equations of Motion (HEOM) approach \cite{HEOM2}, which is rather successful in simulating open quantum systems' dynamics at room-temperature \cite{HEOM1,HEOM3}. However, it exhibits exponential scaling with the size of the system \cite{HEOM2}. On the other hand, the Time Evolving Density operator with Orthogonal Polynomials Algorithm (TEDOPA) technique \cite{TEDOPA1,TEDOPA2,TEDOPA3}, or its more recent extension named Thermalized TEDOPA (T-TEDOPA) \cite{TTEDOPA}, avoids such an intractable scaling as a function of system's size by making use of Matrix Product States (MPS) in simulations of the dynamics of one-dimensional quantum systems \cite{TTEDOPA}. The main bottleneck of this technique however, lies in the scaling of the bond dimension of the MPS, which can increase exponentially as a function of time \cite{MPS1,arealaw}, hence long-time simulations are generally hard to reach via this technique. Simulations of higher-dimensional systems are also difficult to be implemented using MPS, incurring in an exponential scaling of the computational effort as a function of the size of the system \cite{MPS1}. 

One possible alternative approach is to use a quantum computer to simulate open quantum systems~\cite{feynman2018simulating}. Current Noisy-Intermediate Scale Quantum (NISQ) computers are hampered by noisy hardware, low qubit count and time-expensive access. Hence, quantum digital simulations have been applied only to small quantum systems \cite{science,fermihubbard}. In the context of quantum algorithms to simulate open quantum systems, several techniques have recently been proposed, such as using the Kraus operators \cite{KRAUS1,KRAUS2,KRAUS3,KRAUS4,KRAUS5,hu2022general,head2021capturing}, solving the Lindblad equation \cite{STOCHASTIC,cleve2016efficient,VQS,hu2022general}, using the inherent decoherence of the quantum computer to implement the dissipative evolution \cite{decoherencequantumcomputer1,decoherencequantumcomputer2}, discretization of a continuous harmonic oscillator bath \cite{wang2011quantum}, or real-time quantum Monte Carlo algorithms \cite{bauer2016hybrid}, among others \cite{rungger2019dynamical,LIGHT-MATTER, fermion-boson,imaginary,energytransport,georgescu2014quantum}. These techniques face some important issues, namely, the Kraus operators are hard to calculate in practice, Lindblad-form equations are obtained from perturbative expansions of the dynamics generator \cite{breuer2002theory}, hence they are not exact for many realistic systems~\cite{quantumeffectsbook,christensson,profmikhail,polariton,vibelectron}. Regarding the techniques that use the inherent decoherence of the quantum computer to emulate the noise in the simulated system, we point out that the noise is not fully controllable and its structure is not known, hence simulating arbitrary environments via these methods is hard. On the other hand, the referenced method relying on discretization of the bath would require high connectivity between the qubits defining the bath modes and the open system for structured spectral densities \cite{wang2011quantum}, a feature that is not desirable for quantum devices with limited qubit connectivity.

In this work, we propose the implementation of the TEDOPA on a quantum device to simulate non-perturbative dynamics of an open quantum system linearly coupled to a bosonic bath, a method we call {\it Quantum} Time Evolving Density operator with Orthogonal Polynomials Algorithm (Q-TEDOPA) to discern from the original TEDOPA algorithm implemented on a classical computer.  
The Q-TEDOPA is based on a unitary transformation of the environment and system-environment interaction Hamiltonians. In summary, in this work:\\  
(i) We introduce Q-TEDOPA as a quantum simulation technique beyond the collisional models and 
fundamentally different from the existent methods to simulate open system dynamics on quantum devices; Using an orthogonal polynomial transformation of the Hamiltonian, the simulation of an open system linearly coupled to a bosonic bath can be implemented using closed system quantum simulation methods with local qubit interactions, i.e. the transformed environment and system-environment interaction Hamiltonians contain only local qubit interactions. The main differences of Q-TEDOPA in comparison with previously proposed techniques is that the former is well-suited for devices with limited qubit connectivity and well-established quantum simulation methods for closed systems can be used to compute the evolution of an open quantum system.\\
(ii) We explain how this method can be used on a quantum device with restricted qubit connectivity and we demonstrate this by implementing the Q-TEDOPA on a superconducting quantum processor with nearest-neighbour qubit connectivity, namely the IBM-Q device, together with a set of quantum error mitigation techniques that allowed us to obtain more accurate results in a $12$ qubit quantum simulation.\\
(iii) We analyze in detail the complexity and estimate the computational resources of the TEDOPA implemented on a quantum device (Q-TEDOPA) and present arguments that this method can substantially reduce the execution runtime of open quantum systems' simulations over the classical TEDOPA for higher-dimensional systems with loop structures. We also find that the Q-TEDOPA implies no exponential scaling of computational resources in the asymptotic limit of the time length of the simulation (for one-dimensional systems) and system's size (for higher-dimensional systems), a feature that is relevant for fault-tolerant quantum simulation and may potentially yield exponential speedups over classical TEDOPA for some classes of open system models.

The article is organized as follows. In Section \ref{TEDOPA}, we briefly review the TEDOPA technique. In Section \ref{QTEDOPA}, we describe how to implement the TEDOPA on a quantum computer (Q-TEDOPA) and analyze its complexity. Lastly, in Section \ref{numerical}, we apply the Q-TEDOPA to the exciton transport between two light-harvesting molecules employing a quantum processor based on superconductor qubits, provided by the IBM.

\section{Revisiting the TEDOPA}\label{TEDOPA}
Consider an open quantum system composed by some entities (such as molecules) interacting between themselves and with an environment. The total Hamiltonian of such a system can be written as
\begin{equation}
    H =H_{S}+H_{E}+H_{SE}\,, \label{H}
\end{equation}
where $H_{S}$, $H_{E}$ and $H_{SE}$ are the open quantum system, the environment, and the system-environment interaction Hamiltonians, respectively.

We define the environment as a continuum of harmonic oscillators at temperature $\text {T}=0$  and the linear system-environment interaction Hamiltonian reads as follows \cite{TTEDOPA},
\begin{align}\label{H1}
    H_{E} &=\int_{0}^{\infty} d\omega \omega a^{\dag}_{\omega}a_{\omega}, \\
    H_{SE} &= \hat{A} \otimes \int_{0}^{\infty} d\omega \sqrt{J(\omega)} (a^{\dag}_{\omega}+a_{\omega})\,.\label{H2}
\end{align}
The operator $\hat{A}$, which form will be defined later, acts only upon the open quantum system's Hilbert subspace. The operator $a^{\dag}_{\omega}$ ($a_{\omega}$) is the phonon creation (annihilation) operator associated to a mode with frequency $\omega$, such that $[a_{\omega},a^{\dag}_{\omega'}] = \delta(\omega-\omega')$ and $[a_{\omega},a_{\omega}]=[a^{\dag}_{\omega},a^{\dag}_{\omega}]=0$. The spectral density $J(\omega)$ defines the system-environment coupling (squared) times the density of states of the bath \cite{TEDOPA1}. 

We now apply a change of basis transformation to the environment operators as follows \cite{TEDOPA1,TEDOPA2,TTEDOPA}, 
\begin{equation}
    b^{\dag}_{n} = \int_{0}^{\infty} d\omega U_{n}(\omega)a^{\dag}_{\omega}, \quad n = 0,1,\dots \label{transf}
\end{equation}
and to the phonon annihilation operator $a_{\omega}$, yielding $b_{n}$. The resulting operators $b^{\dag}_{n}$ and $b_{n}$ preserve the bosonic commutation relations, i.e. $[b_{n},b^{\dag}_{m}] = \delta_{nm}$ and $[b_{n},b_{m}]=[b^{\dag}_{n},b^{\dag}_{m}]=0$. The transformation $U_{n}(\omega)$ is given by~\cite{TEDOPA1,TEDOPA2}: 
\begin{equation}
    U_{n}(\omega) = \sqrt{J(\omega)}p_{n}(\omega)\,, \label{transform}
\end{equation}
where $p_{n}(\omega)$ is an orthogonal polynomial with respect to the measure $d\mu = J(\omega) d\omega$ defined in the domain $\omega \in [0,\infty)$, i.e. $\int_{0}^{\infty} p_{n}(\omega)p_{m}(\omega)d\mu(\omega) = \delta_{nm}$. 

\begin{figure}
\centering
\includegraphics[width=0.48\textwidth]{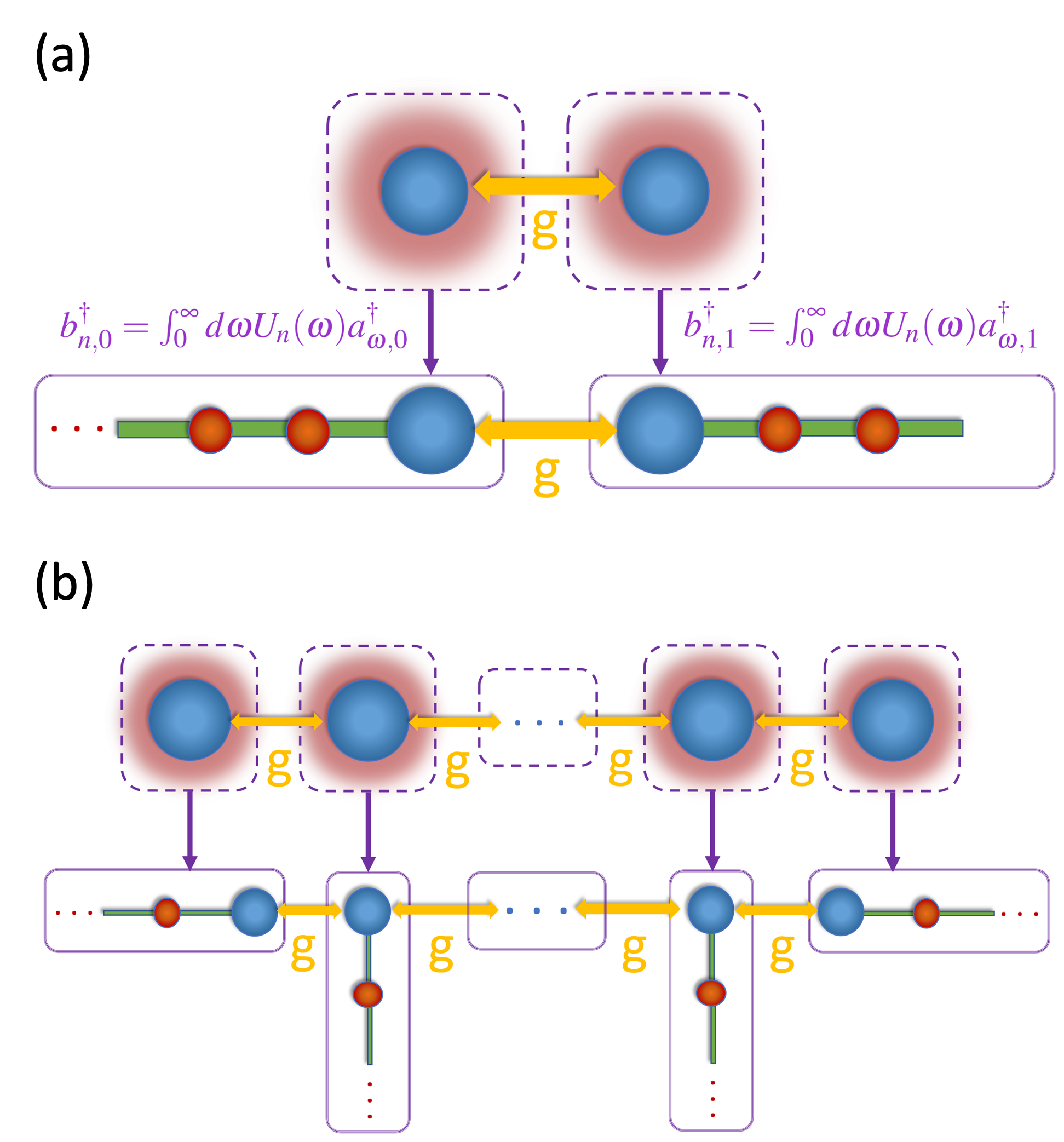}
\caption{(a) TEDOPA Hamiltonian transformation. Initially, the two sites (blue circles) are coupled with interaction strength $g$ and each site interacts with its own identical harmonic oscillator environment (red blur surrounding the sites). The TEDOPA transformation is applied to each environment and site-environment interaction Hamiltonians, i.e. to the environmental operators $a^{\dag}_{\omega}$ and $a_{\omega}$, such that the resultant model comprises a nearest-neighbour semi-infinite one-dimensional system representing the environment. On the left (right) edge of the one-dimensional system, one has a chain of harmonic oscillators coupled to the first (second) site. This chain represents the environment, where each red circle is a harmonic oscillator. (b) Same as (a), but now with a higher number of coupled sites. A cacti-type network is obtained after applying the TEDOPA Hamiltonian transformation, hence MPS-based simulations are no longer so efficient as in the case (a), as explained in the main text.}
\label{fig:chainall}
\end{figure}

Using the orthogonal polynomial recurrence relation \cite{TEDOPA2} and the definition of an orthogonal polynomial presented above, one can exactly map the environment and interaction Hamiltonians to the chain Hamiltonian as follows,
\begin{align}
    H^{C} &= H^{C}_{SE}+H_{E}^{C}, \\
    H^{C}_{SE} &= t_{0}\hat{A}\otimes (b^{\dag}_{0}+b_{0})\,,\label{Hc1} \\
    H_{E}^{C} &= \sum_{n=0}^{\infty}w_{n}b^{\dag}_{n}b_{n}+\sum_{n=1}^{\infty}t_{n,n-1}b^{\dag}_{n}b_{n-1}+h.c. \, ,\label{Hc2}
\end{align}
\noindent where the Hamiltonian parameters $t_{0}$, $w_{n}$ and $t_{n,n-1}$ can be obtained from the orthogonal polynomial recurrence coefficients. The TEDOPA Hamiltonian transformation is illustrated in Fig. \ref{fig:chainall} for two- and $n$-site coupled open quantum systems.

In order to map the Hamiltonians given by Eqs. \eqref{H1} and \eqref{H2} to the chain Hamiltonian, the recurrence coefficients of the orthogonal polynomials defined with respect to the measure $d\mu(\omega)=J(\omega)d\omega$ must be calculated. They can be obtained analytically for strictly Ohmic baths \cite{TEDOPA2}, or, in general, through a numerically efficient (classical) algorithm, available in the \textit{ORTHPOL} package \cite{ORTHPOL}.

Some remarks concerning the exact TEDOPA transformation are in order: (\textit{i}) commutation and anti-commutation relations of the initial bath operators are preserved after the transformation. This means that one can also simulate, for instance, fermionic baths.
(\textit{ii}) If we define a spectral density belonging to the Szegö class \cite{TEDOPA5} with hard frequency cutoffs $\omega_{min}$ and $\omega_{max}$, the chain Hamiltonian coefficients converge \cite{TEDOPA4,TEDOPA5} to
\begin{equation}
\lim_{n\to \infty} w_{n}=\frac{\omega_{max}+\omega_{min}}{2}\,,\label{limit1}
\end{equation}
and 
\begin{equation}
\lim_{n\to \infty} t_{n,n-1}=\frac{\omega_{max}-\omega_{min}}{4}.\,\label{limit2}
\end{equation}
(\textit{iii}) A generalization of TEDOPA to nonzero temperature environments has been termed Thermalized TEDOPA (T-TEDOPA) \cite{TTEDOPA}. It consists in defining a temperature-dependent spectral density \cite{TTEDOPA} as follows,
\begin{equation}
    J_{\beta}(\omega) = sign(\omega)\frac{J(|\omega|)}{2}\left(1+\coth(\frac{\beta \omega}{2})\right). \label{jbeta}
\end{equation}

The T-TEDOPA transformation, in contrast with equation \eqref{transf}, is given by:
\begin{equation}
    b^{\dag}_{\beta,n} = \int_{-\infty}^{\infty} d\omega U_{\beta,n}(\omega)a^{\dag}_{\omega}, \quad n = 0,1,\dots \,,
\end{equation}

where $\beta ^{-1}={k_{B}\text{T}}$ and $U_{\beta,n}(\omega)=\sqrt{J_{\beta}(\omega)}p_{\beta,n}(\omega)$, such that the temperature-dependent orthogonal polynomials are defined with respect to the measure $d\mu_{\beta}(\omega)=J_{\beta}(\omega)d\omega$. The initial state of the harmonic oscillators in the chain is always set to the vacuum state, following reference \cite{TTEDOPA}.
\label{sec:develop}

\section{Quantum TEDOPA} \label{QTEDOPA}

\subsection{Implementation}
The basic insight is that the evolution operator of the transformed Hamiltonian $H^{C}$ can be efficiently implemented on a quantum computer to simulate the environment. The simulation of the action of the environment on the open system is where the main bottleneck of simulations of non-perturbative dynamics of open quantum systems lies because the continuous (infinite-dimensional) description of the environment is intractable if not appropriately manipulated, e.g. with the TEDOPA transformation. Therefore, herein, we restrict ourselves by assessing the complexity of simulating the environment Hamiltonian with Q-TEDOPA.
In order to numerically simulate the quantum system, the chain in Q-TEDOPA must be truncated to some number, $l$, of harmonic oscillators in it, and the Hilbert space of each harmonic oscillator must be truncated to a $d$-dimensional site (allowing up to $(d-1)$ phonons for a site $n$). Afterwards, the environment creation and annihilation operators obtained in equation \eqref{transf} must be mapped to Pauli operators so that the chain of harmonic oscillators is appropriately encoded in a quantum circuit. As extensively studied in reference \cite{bathoperators}, choosing a particular bosonic qubit encoding may change how the number of necessary quantum gates and qubits to simulate the system scale in the asymptotic limit of $l$ and $d$ tending to infinity. In Appendix \ref{sec:bosonic}, we describe how to map bosonic operators to Pauli operators using the unary and binary qubit encodings and the scaling of these transformations as a function of the dimensions of the harmonic oscillator chain.

\begin{figure*}[t]
\centering
\includegraphics[width=1\textwidth]{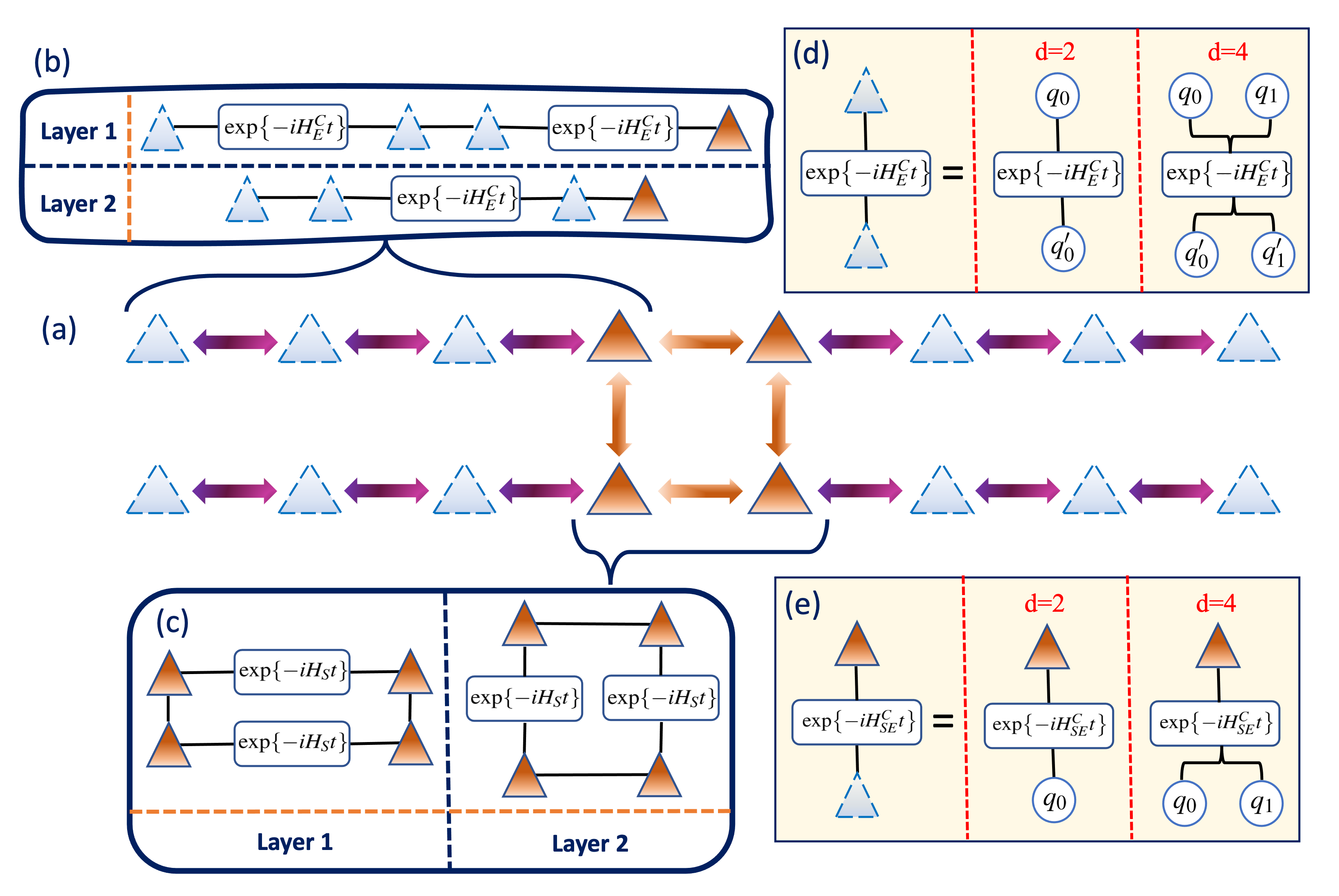}
\caption{Implementation of Q-TEDOPA. After applying the TEDOPA Hamiltonian transformation, one obtains a nearest-neighbour interaction model as shown in (a). (a) Open quantum system composed of $4$ two-level molecules (orange triangles) with nearest-neighbour interactions as illustrated by the orange arrows. Each molecule interacts with a chain truncated to three harmonic oscillators (represented by blue triangles). Purple arrows represent the two-site interactions in the chain Hamiltonian $H^{C}$. Inside the boxes (b) and (c), we show the types of two-qubit gate layers that are encoded on the quantum circuit to simulate the evolution of the total system via the Trotter-Suzuki product formula. These are illustrated for the chain Hamiltonian in the box (b) and for the open quantum system Hamiltonian in the box (c). (d) Example of the implementation of the interaction between two harmonic oscillators, mapped to sets of qubits $q$ and $q'$ (represented by the blue circles). We illustrate the implementation of this interaction on a quantum circuit for $d=2$ and $d=4$ binary qubit encodings of the harmonic oscillators (more details in Appendix \ref{sec:bosonic}). (e) Same as (d) but for the molecule-harmonic-oscillator interaction.} 
\label{fig:circuit_structure}
\end{figure*}

Having the chain Hamiltonian written in terms of Pauli operators as discussed in Appendix \ref{sec:bosonic}, one can perform the time evolution with an efficient quantum algorithm that performs closed Hamiltonian dynamics simulations. The TEDOPA transformation allows one to solve the problem of simulating an open system by simulating a closed system instead. This is specially relevant for quantum computing because the quantum gates available to perform the computation are unitary. To simulate the system on a quantum computer, the evolution operator may be decomposed with Trotter-Suzuki decompositions \cite{TROTTER1,TROTTER2}, truncated Taylor series \cite{TAYLOR} or qubitization and quantum signal processing techniques \cite{qubitization}. The simulations performed with these decomposition techniques typically require long coherence times, that
for current NISQ computers can produce unfaithful results. Therefore, one may instead simulate the closed Hamiltonian dynamics via hybrid quantum-classical methods, such as the Variational Quantum Simulator \cite{VQS, VQS2}, Variational Fast-Forwarding methods \cite{VFF,VHD} or variational compressed Trotter techniques \cite{VQD,VQD2}. 

We remark that the chain Hamiltonian is local, in the sense that it exhibits only nearest-neighbour interactions in the environment Hilbert subspace. On quantum hardware, a $d$-dimensional harmonic oscillator is encoded in a set of qubits. For $d=2$, only one qubit per harmonic oscillator is required. This means that the resulting quantum circuit to simulate the dynamics exhibits solely nearest-neighbour qubit-qubit interactions as shown in Figure \ref{fig:circuit_structure}. For $d$-dimensional harmonic oscillators, where $d>2$, the range of qubit-qubit interactions is, in the worst case, $2d$-local for unary encoding and $2\log(d)$-local for binary encoding. Hence, for $d=1,2$, Q-TEDOPA can be already implemented on quantum hardware with limited qubit connectivity such as superconducting quantum processors. A larger number of phonons $d$ requires the implementation of SWAP operators to apply the two-oscillator interactions in the chain. This yields a higher average circuit error rate which may produce unfaithful simulation results in current quantum computers. However, given the fast development of quantum devices in the past few years, we expect that the increase of two-qubit gate fidelities in the future will alleviate this issue since the non-locality, i.e., the logarithmic (linear) scaling of $H^{C}$ as a function of the number of phonons $d$ for binary (unary) encoding, will not pose a bottleneck for the simulation of Q-TEDOPA (we recall that very large $d$, chosen typically to be in the interval $[10,100]$, e.g. see Appendix \ref{app: resource_est}, are usually not required for room-temperature calculations). In Figure \ref{fig:circuit_structure}, we illustrate how the Q-TEDOPA can be implemented for a four-site open quantum system, where each site is coupled to a harmonic oscillator environment. 

\subsection{Errors in the Q-TEDOPA}

Two sources of error arise in Q-TEDOPA simulations: chain truncation error due to the approximation of the TEDOPA transformation with a finite number of orthogonal polynomials up to a simulated time $t$, and the approximated evolution method one chooses to evolve the transformed open quantum system.
The restricted number of phonon states per site, $d$, depends on the structure of the spectral density, i.e. it is specific for a given problem, as investigated in reference \cite{TEDOPA4}. Typically, it is not required to be very large, even for room-temperature calculations as shown in references \cite{TEDOPA4,TTEDOPA} (e.g. for the temperature $\text {T}=300$~K, $d=12$ provides accurate results for a realistic photosynthetic environment). As shown in reference \cite{woods2015simulating}, a rigorous upper bound on the simulation error arising from truncating the number of environmental phonon states in a generalized spin-boson model has been calculated, showing that this is suppressed exponentially for large $d$.

An heuristic estimate has been proposed~\cite{TEDOPA4} to evaluate the chain length as $l=2t_{\infty}T$, where $T$ is the maximum simulated time and $t_{\infty}=\lim_{n\to \infty} t_{n,n-1}$ the chain hopping term, as defined in equation \eqref{limit2}. Furthermore, empirical investigations \cite{TEDOPA4,nusseler2022fingerprint} suggest that the linear scaling of $l$ as a function of the maximum simulated time $T$ provides accurate results. Reference \cite{woods2015simulating} provides an upper bound of the error of a measured observable for the open system. This bound decays superexponentially as a function of $l$ as long as the open system dynamics is defined outside of a Lieb-Robinson-type light-cone $cT<l$. Nevertheless, the error of the measured observable can be calculated \textit{a priori}, for instance, by using the work reported in reference \cite{mascherpa2017open}. Herein, the error can be inferred by computing the deviations between the spectral densities that characterize harmonic oscillator chains with different truncated numbers of oscillators $l$.

We note that the total error of the dimensional truncation of the chain is the sum of the errors that arise from truncating $d$ and $l$ \cite{woods2015simulating}, hence the chain truncation procedure does not theoretically pose a challenge for simulations of non-perturbative dynamics of systems linearly coupled to bosonic environments with the Q-TEDOPA.

A possible procedure to find the minimum $l$ and $d$, though variational, is to simulate the system for a time $t$ with an initial guess of $l$ $d$-dimensional harmonic oscillators, measure a local observable $\hat{O}$ on the open quantum system Hilbert subspace, and then repeat the simulation procedure for the same time, with $l'\geq l$ of $d'-$dimensional harmonic oscillators with $d' \geq d$. Choosing a error threshold $\varepsilon>0$, numerically exact results are obtained if $|\langle \hat{O} \rangle_{t}-\langle \hat{O}' \rangle_{t}|<\varepsilon$, where $\hat{O}'$ is the measured observable in the simulation with the increased chain dimensions $d'$ and $l'$. On the other hand, other error metrics may be employed such as the \textit{total variation distance} \cite{randcomp} or the fidelity \cite{nielsenchuang} between the simulations. In the former (latter), one can minimize the absolute difference between the probability distributions of the qubits in the open system (maximize the overlap between the states of the qubits in the open system) obtained by executed simulations with different truncated number of oscillators. Note that these heuristic estimates in Q-TEDOPA depend only on the number of qubits in the open system.

The latter source of error in the Q-TEDOPA comes from the approximated evolution method one chooses to simulate the closed quantum dynamics of the transformed open quantum system. Herein, we focus on the first-order Trotter-Suzuki product formula decomposition of the evolution operator, which we use as the evolution method to simulate an open quantum system with the Q-TEDOPA on an IBM-Q device in Section \ref{numerical}. The reason for this choice is that this technique has been suggested to be the most efficient evolution method to simulate closed quantum system dynamics on NISQ devices \cite{childs2018toward}. The algorithmic error of such an evolution method scales with the oscillator chain size as \cite{childs2021theory},
\begin{equation}
    \epsilon = O(\alpha_{comm}T^2/N)\leq O(l\left[d^2log(d)T\right]^2/N),
\end{equation} 
with $T$ being the maximum simulation time, $N$ the number of Trotter iterations and $\alpha_{comm}=\sum_{\gamma_{1},\gamma_{2}}||[H_{\gamma_{2}},H_{\gamma_{1}}]||\leq O(l[d^2log(d)]^2)$ (see Appendix \ref{order_comm}) for the Hamiltonian $H^C$ decomposed in Pauli strings (with binary encoding), where $||\cdot||$ denotes the operator norm. Higher $k$-th order Trotter-Suzuki decompositions can be also employed to exponentially decrease the algorithmic error with $k$, however the number of quantum gates will increase exponentially as a function of $k$ \cite{wiebe2010higher}, thus the hardware error will quickly dominate over the algorithmic error on the current NISQ era devices. We also refer the reader to recent advances in reduction of the algorithmic error of product formulas which can be further exploited in the Q-TEDOPA. These include symmetry-protected formulas \cite{tran2021faster}, random formulas \cite{chen2021concentration} or implementation of a specific Trotterisation sequence of Hamiltonian terms that preserves the locality of the simulated system \cite{childs2021theory}.

\subsection{Comparison of the complexity of the Q-TEDOPA and TEDOPA}\label{sec:develop2}

\begin{figure*}[t!]
\centering
\includegraphics[width=0.95\textwidth]{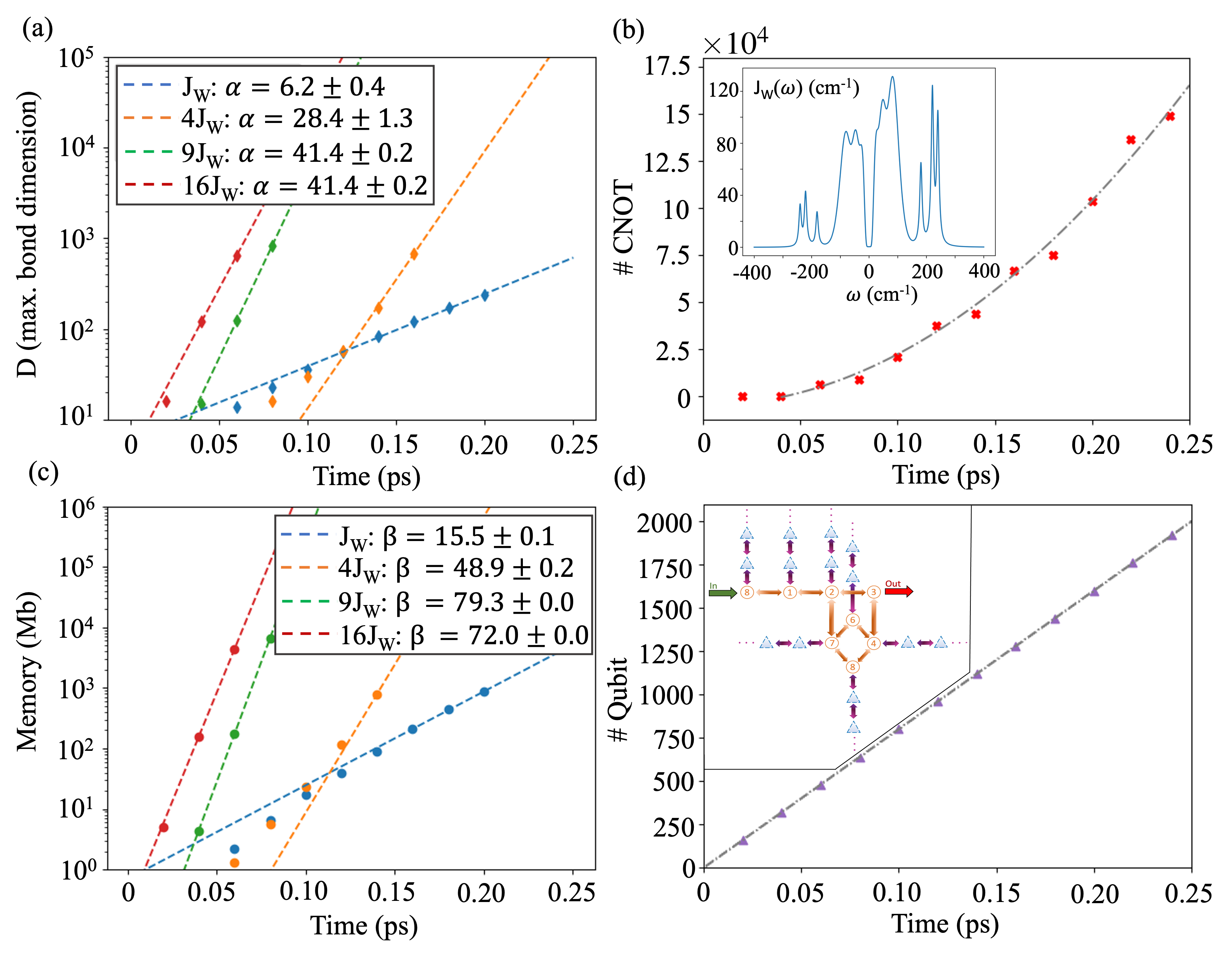}
\caption{Scaling of the computational resources in the classical T-TEDOPA, (a) and (c), and in the Q-TEDOPA, (b) and (d), for the simulation referred in the main text. The scatter plot shows the results obtained, whereas dashed lines represent fitted scaling functions to the results, i.e. (a) $O(e^{\alpha t})$, (b) $O(t^2)$, (c) $O(e^{\beta t})$ and (d) $O(t)$, where $t$ denotes time. (a) Scaling of the maximum MPS bond dimension obtained with the classical T-TEDOPA simulation with MPS truncation error $\varepsilon<10^{-8}$ for several system-environment couplings, i.e. magnitude of $J_{W}(\omega)$. Note that the execution runtime of the T-TEDOPA scales as $O(D^3)$ (neglecting other minor multiplicative terms). (b) Number of required CNOT gates in the Q-TEDOPA to implement the quantum digital simulation of the same system. Inset shows the spectral density $J_{W}(\omega)$, consisting of a broad background plus three Lorentzian peaks, at T$=300$~K. (c) Memory used in the classical T-TEDOPA simulation to store the set of MPS that describe the total system. (d) Number of qubits required to simulate the same open system using the Q-TEDOPA. Inset shows the qubit encoding of the FMO complex used in the quantum digital simulation (see Fig. \ref{fig:circuit_structure} for the description of each element). In both simulations, $d=16$ (binary qubit encoding was employed in the Q-TEDOPA) and the $2^{nd}$ order Trotter-Suzuki product formula was used (with the same Trotter time-step). The employed Hamiltonian and parameters are reported in Section \ref{numerical} and Appendix \ref{sec:coeffs}.1, respectively. The missing reorganization energy ratio of the considered spectral density is $5.75\times 10^{-4}$ (see Appendix \ref{sec:coeffs}.2).
}
\label{fig:scaling}
\end{figure*}

Simulations of Hamiltonian dynamics with the TEDOPA on a classical computer are usually performed using  MPS-based techniques for one-dimensional systems, such as the Time-Evolving Block Decimation (TEBD) algorithm \cite{TTEDOPA, MPS1}. For instance, the execution time of the TEBD algorithm in combination with the TEDOPA has been reported to asymptotically scale as $O(lN(d D)^3)$ \cite{TTEDOPA}. We denote by $N$ the number of Trotter iterations and by $D$ the bond dimension of the MPS. The absence of an upper bound for the entanglement entropy as a function of time for non-equilibrium systems' simulations opens up the possibility of having the bond dimension exponentially scaling with time, i.e. $D \sim 2^{t}$, such as in the presence of quantum quenches \cite{arealaw}, hence limiting classical simulations to short time lengths (see Fig. \ref{fig:scaling}(a)). Furthermore, in higher-dimensional systems, the problem persists and the bond dimension additionally incurs in an exponential scaling with the size of the system \cite{MPS1}. For instance, a simulation of the exciton transport across a chain of several light-harvesting molecules would be mapped into a two-dimensional system after applying the TEDOPA Hamiltonian transformation, as shown in Figure \ref{fig:chainall}b. Therefore, the system for a large number of molecules is not efficiently simulated with MPS \cite{dunnett2021influence}, as the bond dimension would blow up exponentially with the size of the harmonic oscillator chain \cite{MPS1}. Recently, a set of techniques has been reported to efficiently simulate quasi- one-dimensional systems such as the one in Figure \ref{fig:chainall}b for a small number of molecules, through the use of tree-MPS \cite{dunnett2021influence}. Other techniques to simulate TEDOPA rely on Variational Matrix Product States (VMPS) and other sophisticated methods \cite{schroder2016simulating,del2018tensor}, however the crux of the problem remains: for large open quantum systems, i.e. a chain of a large number of coupled molecules in interaction with the environment as shown in Figure \ref{fig:chainall}b, or system's topologies with loops, e.g. the one in Figure \ref{fig:circuit_structure}, MPS-based methods are not efficient.

Regarding the implementation of the TEDOPA on a quantum computer, i.e. the Quantum TEDOPA, by choosing a first-order Trotter-Suzuki decomposition of the evolution operator as the time evolution method, one may simulate the dynamics with the number of quantum gates asymptotically scaling as $O(lN d^2)$ for unary qubit encoding and $O(lN d^2 \log(d))$ for binary qubit encoding (see Appendix \ref{sec:bosonic}). The number of qubits required to encode the chain in the simulation is $ld$ for unary qubit encoding and $l\log(d)$ for binary qubit encoding. Given the available noisy quantum hardware, one can trade a low qubit count for a higher circuit depth by adopting binary qubit encoding or the opposite with the choice of unary qubit encoding. The number of necessary quantum gates and qubits, i.e. computational resources, does not grow exponentially with the time length of the simulation or size of the system for the Q-TEDOPA, and this efficient behaviour is kept for simulations of higher-dimensional systems. A Q-TEDOPA simulation of a two-dimensional system performed with a Trotter-Suzuki product formula is exemplified in Figure \ref{fig:circuit_structure}.

In Fig. \ref{fig:scaling}, we compare the runtime execution and memory required to run the T-TEDOPA on a classical and quantum computer (Q-TEDOPA) and measure the population terms of the open system's density matrix (in the quantum computer this is equivalent to measure the computational basis states in all open system qubits). We estimate that a classical T-TEDOPA simulation of the exciton transport between two molecules, each coupled to a characteristic, structured photosynthetic spectral density $J_{W}(\omega)$ \cite{TTEDOPA,chandrasekaran2015influence} at temperature $\text {T}= 300$~K, shown in the inset of Fig. \ref{fig:scaling}(b), yields an exponential increase of the execution runtime and memory on a classical computer as demonstrated in Figs. \ref{fig:scaling}(a) and \ref{fig:scaling}(c), respectively. In general, by increasing the system-environment coupling in a classical T-TEDOPA simulation, the execution runtime and memory are also increased as shown in Figs. \ref{fig:scaling}(a) and \ref{fig:scaling}(c). This is due to the stronger interaction between the open system and the harmonic oscillator chains, which produce a higher degree of correlations between these, resulting in higher bond dimensions in the MPS representation of the total system. On the other hand, in a simulation with the Q-TEDOPA, the change of this coupling does not increase the amount of necessary computational resources to be used in a quantum computer, i.e. CNOT gate or qubit count, relatively to the classical simulation with T-TEDOPA, because the total state of the quantum system is fully represented in the qubits. This is demonstrated in Figs. \ref{fig:scaling}(b) and \ref{fig:scaling}(d), where we display estimates of the quantum runtime execution and quantum memory, i.e. CNOT gate and qubit counts, respectively, as a function of the time for the simulation of the same system using the Q-TEDOPA. We observe a quadratic and a linear scaling with time $T$ for CNOT gate and qubit counts, respectively, which is due to the increase of the number of Trotter iterations $N=T/\Delta t$ and in accordance with the chain length heuristic estimate $l=2t_{\infty}T$ used here (the former is true only for the CNOT gate count). 

We also quantify the computational resources of a simulation of the exciton transport in a widely studied photosynthetic system, the Fenna-Matthews-Olson (FMO) complex of green sulphur bacteria, which contains $8$ molecules in a monomer, each coupled to a harmonic oscillator bath \cite{ishizaki2009theoretical,thyrhaug2018identification}. We encode the molecules of the FMO complex and the strongest interactions among them in a 2D quantum circuit architecture, such as the one present in the quantum processor Google Sycamore \cite{harrigan2021quantum}. We then transform the Hamiltonian, following the TEDOPA transformation presented in Section \ref{TEDOPA}, yielding a chain of harmonic oscillators coupled to each molecule. The resulting quantum circuit structure is shown in the inset of Fig. \ref{fig:scaling}(d). We predict that a Q-TEDOPA simulation, where each molecule is coupled to a bath defined by the spectral density shown in the inset of Fig. \ref{fig:scaling}(b), for a simulation up to $T=0.5$ $ps$, requires $488$ qubits and $\sim 2.88 \times 10^6$ CNOT gates, assuming no circuit optimizations, no optimized reduction of oscillators' energy levels $d$ (see \cite{TTEDOPA}) and neglecting SWAP operators. On the other hand, using the estimates from Fig. \ref{fig:scaling}(a) and \ref{fig:scaling}(c), a classical T-TEDOPA simulation of only $2$ molecules in the FMO complex for $T=0.5$ $ps$ would require $D \sim 6.05\times 10^{4}$ and total memory $38.5$ Tb to achieve an error rank truncation error $\varepsilon<10^{-8}$. In Appendix \ref{app: resource_est}, we calculate the number of CNOT gates and qubits required to execute previously reported models with state-of-the-art tensor network simulations (via T-TEDOPA) \cite{chin2013role,dunnett2021influence} on quantum computers (with Q-TEDOPA). These resource estimates suggest that Q-TEDOPA may be appropriate to be employed when low-medium temperature environments are desirable to be simulated, since large $d$ have a notorious impact on the quantum simulation (see Appendix \ref{app: resource_est}). The classical simulations \cite{chin2013role,dunnett2021influence} considered in these estimates contain only a few open system's degrees of freedom (these do not exceed two sites) due to the difficulty of tensor networks to simulate large, higher-dimensional systems. Therefore, Q-TEDOPA may yield an advantage over the classical TEDOPA when the open system model desired to be simulated contains more than a few units of degrees of freedom arranged on a two-dimensional structure with loops and the bath temperature is not high.

\subsection{Q-TEDOPA on near-term quantum hardware}
Recently, quantum simulations of medium-large quantum systems \cite{kim2023evidence,van2023probabilistic,dborin2022simulating} have been performed on superconducting quantum devices, demonstrating that more complex simulations than the ones realized in this work (see Section \ref{numerical}) are possible on near-term quantum devices by optimizing the quantum circuits and implementing well-suited quantum error mitigation techniques. The ability to run such experiments on near-term quantum hardware, together with the complexity analysis and simulations reported in this work, suggest that current superconductor quantum devices may be able to perform Quantum TEDOPA simulations of systems containing oscillators with the number of energy levels up to $d = 2$ (i.e. $1$ qubit encoding an harmonic oscillator), small to medium-sized oscillator chains ($l<20$) and open system degrees of freedom encoded in a number in the order of units of qubits for a small number of Trotter iterations, i.e. short time periods. However, we note that an appropriate estimate of $d$, $l$, the number of necessary Trotter iterations required to simulate the open system up to a time $t$, and the number of qubits encoding the open system degrees of freedom must be obtained in order to infer if the simulation is feasible on current quantum hardware. For instance, the quantum device may simulate a higher number of open system sites encoded in the qubits at the cost of a reduced number of energy levels, $d$, per oscillator or reduced chains size, $l$, hence the system that can be simulated on the quantum hardware is highly dependent on the trade-off of these computational resources. Since classical tensor network methods can tackle large chain sizes, $l$, with hundreds of oscillators with encoded energy levels well above $d = 2$, but only a few coupled open system sites encoded in the simulation \cite{dunnett2021influence,schroder2016simulating,del2018tensor}, the Quantum TEDOPA may be useful on the near-term quantum hardware for two-dimensional open quantum systems with a small-medium number ($\sim 8-10$) of coupled sites structured in a loop architecture that can be simulated by current circuit architectures of the IBM and Google quantum devices (e.g. see inset of Fig. \ref{fig:scaling}(d)). A proper assessment of the computational resources of simulations of quantum systems that satisfy these requirements will be done in a future work.

\section{Numerical implementation} \label{numerical}
We simulated the exciton transport between two electromagnetically coupled molecules of the photosynthetic FMO complex~\cite{quantumeffectsbook} in a proof-of-concept scenario. The Hamiltonian of the whole system is given by equation \eqref{H}. The open quantum system is composed of just two molecules and its Hamiltonian reads:
\begin{equation}
    H_{S}=\sum_{m=0}^{1}\epsilon_{m} c^{\dag}_{m}c_{m}+ g\sum_{n\neq m}c^{\dag}_{m}c_{n}, \label{HS}
\end{equation}
where $\epsilon_{m}$ is the energy of the excited state of molecule $m$, $g$ is the electromagnetic coupling between the molecules, and $c^{\dag}_{m}$ ($c_{m}$) is the exciton creation (annihilation) operator applied to the molecule $m$. Appendix \ref{sec:coeffs} reports the values of the Hamiltonian parameters. We map the exciton operators to Pauli operators using the hardcore boson qubit encoding as explained in Appendix \ref{C}.

\begin{figure*}
\centering
\includegraphics[width=\textwidth]{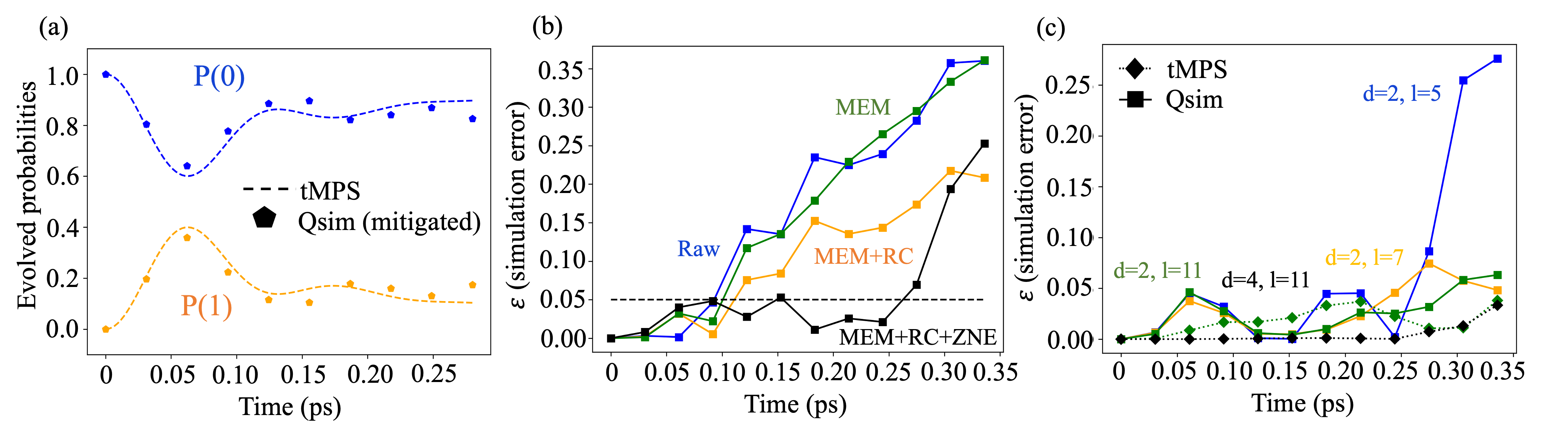}
\caption{(a) Error-mitigated results of the quantum digital simulation performed on the \textit{ibmq toronto} (pentagons) and numerical results obtained with tMPS (dashed lines). $P(0)$ ($P(1)$) denotes the probability of the exciton to be in the molecule $0$ ($1$). (b) Error of the quantum simulation executed on the \textit{ibmq toronto} when performed with and without quantum error mitigation techniques: \textit{Raw}, \textit{MEM}, \textit{RC} and \textit{ZNE} correspond to results obtained without error mitigation and with the measurement error mitigation, randomized compiling and zero-noise extrapolation, respectively. The black dashed line corresponds to the upper bound of the error of the quantum simulation results (up to the $9$th iteration). (c) Error of the classical simulations of quantum circuits without noise (\textit{Qsim}) for different chain lengths. We also show the error of the results of the tMPS simulations (dotted lines) with chains of $11$ two-dimensional and four-dimensional encoded harmonic oscillators (the remaining parameters of the simulation are reported in Appendix \ref{D}).}
\label{fig:results}
\end{figure*}

We choose the $\hat{A}$ operator in the interaction Hamiltonian \eqref{H2} to be the Pauli operator $\hat{Z}$ \cite{nonmarkovian,vibENAQT}, so that each molecule is coupled to an equivalent structured bath as follows, $$H_{SE} = \sum_{m}\hat{Z}_{m} \otimes \int_{0}^{\infty} d\omega \sqrt{J(\omega)} (a^{\dag}_{\omega,m}+a_{\omega,m})\, .$$ The environment Hamiltonian is the one of equation \eqref{H1}. The bath is characterized by an exponentially decaying Ohmic spectral density at $\text {T}=0$~K of the form \cite{TEDOPA2},
\begin{equation}
    J(\omega) = 2\pi \alpha \omega e^{-\omega/\omega_{c}}\theta(\omega), \label{SD}
\end{equation}
where $\alpha=0.25$ is a dimensionless exciton-phonon coupling, $\omega_{c}=100$ $cm^{-1}$ is a soft cutoff frequency and $\theta(\omega)$ is the Heaviside step function. The hard cutoff frequency is chosen as $\omega_{max}=10\, \omega_{c}$ in order to have a negligible influence on the dynamics of the open quantum system (see Appendix \ref{sec:coeffs} for more details). The open quantum system's evolution timescale is set by $\Delta E=E_{\max}-E_{\min}$, where $E_{\max}$ ($E_{\min}$) is the largest (smallest) energy eigenvalue of $H_{S}$ in the single-excitation sector. We remark that the chosen spectral density characterizes a non-Markovian environment, i.e. $J(\pm\Delta E) \approx const$ \cite{nonmarkovian} and it exhibits a moderate open system-environment coupling, i.e. $J(\Delta E) \sim g/2$. The environment and interaction Hamiltonians defined above are transformed to harmonic oscillator chains following the TEDOPA Hamiltonian transformation as described in Section \ref{TEDOPA}.

Figure \ref{fig:chainall}~a illustrates the simulated exciton transport model, where each site in the open quantum system is represented by a qubit, so that $\ket{0}$ and $\ket{1}$ represent the ground and excited states of a molecule, respectively. The phonon operators are mapped to the Pauli operators following a $d=2$ (two energy levels) binary qubit encoding, hence each harmonic oscillator is represented by a single qubit. A total of $5$ harmonic oscillators in each chain were simulated. The initial state of the system is the environment in the vacuum state ($\ket{0}$ for all harmonic oscillator qubits) and the molecular qubits are initialized in state $\ket{10}$, i.e. the molecule $0$ in the excited state and the molecule $1$ in the ground state. The evolution of the open quantum system and the double chain of harmonic oscillators is performed using the first order Trotter-Suzuki product formula. The quantum digital simulation comprises a total of $12$ qubits and about $400$ implemented CNOT gates for $10$ Trotter iterations. In order to obtain reliable results using the quantum hardware, several quantum error mitigation techniques have been implemented, namely, \textit{Qiskit} measurement error mitigation \cite{qiskit}, Randomized Compiling \cite{randcomp,randcomp2} and Zero-Noise (exponential) Extrapolation \cite{zne2,zne3,zne4} via the \textit{Mitiq} package \cite{mitiq}. We used \textit{Qiskit Runtime} to run the quantum simulation and the \textit{ibmq toronto} quantum computer provided by \textit{IBM} as our testbed. Lastly, we measured the probability of obtaining the states $\ket{10}$ and $\ket{01}$ in the molecular qubits, making use of the symmetry verification mitigation technique \cite{mitigationreview}. For more details on the implementation of the quantum simulation, the reader can refer to Appendix \ref{C}.

The simulation results for $10$ Trotter iterations are shown in Figure \ref{fig:results} (and those with 12 iterations are provided in Appendix \ref{C}, Figure \ref{fig:withoutzne}). We compared the quantum simulation results with the results obtained from a numerically exact tMPS simulation \cite{MPS1} (see Appendix \ref{D} for more details). The quantum simulation results agree well with the classically computed ones up to the $9$th Trotter iteration, showing that the current quantum computers can already simulate non-perturbative dynamics of open quantum systems for short time lengths. In Figure \ref{fig:results}b, we plot the error of the quantum simulation when performed with and without the quantum error mitigation techniques. We define the simulation error as $\varepsilon=|\langle\hat{O}\rangle^{(circ)}_{t}-\langle\hat{O}\rangle^{(num. exact)}_{t}| \in [0,1]$ for $\langle\hat{O}\rangle_{t}=\bra{\Psi(0)}e^{iHt}\ket{01}\bra{01} e^{-iHt}\ket{\Psi(0)}$, where $\ket{\Psi(0)}$ is the initial state of all the qubits in the system, and $\langle\hat{O}\rangle^{(circ)}_{t}$ and $\langle\hat{O}\rangle^{(num. exact)}_{t}$ are the measured expectation values in the quantum circuit and numerically exact tMPS simulations, respectively. This error metric is equivalent to the \emph{total variation distance} (TVD), up to a multiplicative factor of $2$ \cite{randcomp}. We observed that the error of the results of the quantum mitigated simulation shown in Figure \ref{fig:results}a is $\varepsilon  < 0.05$ up to the $9$~th Trotter iteration. If Zero Noise Extrapolation (ZNE) is not implemented, the inherent decoherence of the quantum computer causes the probabilities to quickly decay, as demonstrated by the increasing simulation error in Figure \ref{fig:results}b and further discussed in Appendix \ref{C}. On the other hand, when ZNE is implemented, we observe that the probabilities decay slower as demonstrated in Appendix \ref{C}, Figure \ref{fig:withoutzne}. This slow decay is an effect produced by the abrupt truncation of the harmonic oscillator chain, and not by the inherent noise in the quantum computer. The cause of this effect can also be understood by performing a classical simulation without noise of the quantum circuits and comparing its simulation error with the one from the tMPS simulation, as we show in Figure \ref{fig:results}c. As illustrated, increasing the number of encoded oscillators in the quantum circuit from $l=5$ to $l = 7$, the error of the quantum circuit simulation is significantly reduced for long simulation times. However, further increase of $l$ does not result in accuracy improvements. Moreover, the choice of using higher-dimensional harmonic oscillators in the simulation does not appreciably impact the accuracy of the results as shown by the error of the tMPS simulations (relatively to the numerical exact one) executed with different dimensions $d$ in Figure \ref{fig:results}c.

\begin{figure}[t]
\centering
\includegraphics[width=0.48\textwidth]{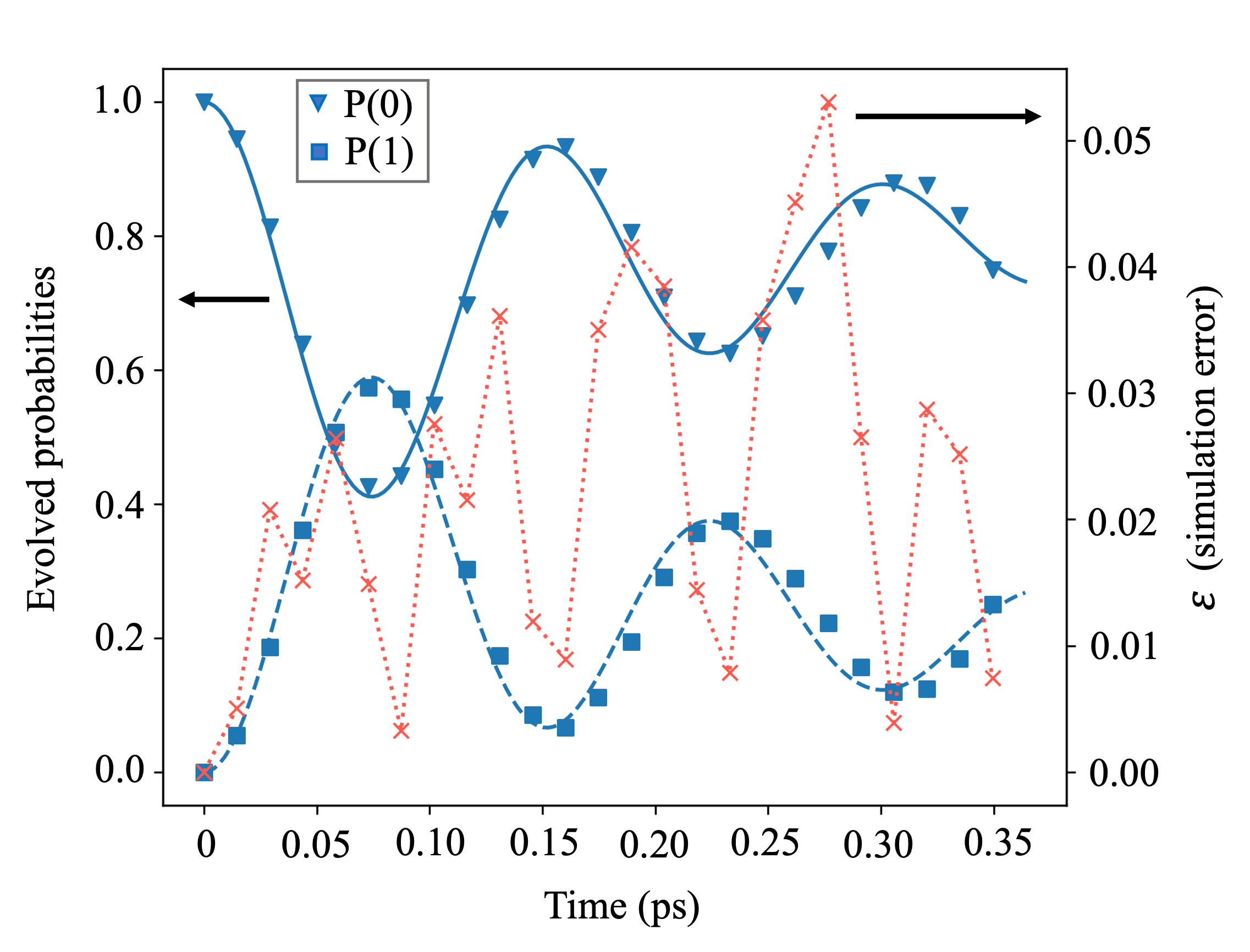}
\caption{Results of the simulation of a nonzero temperature environment using the Q-TEDOPA. We compare the classical simulation of the quantum circuits, i.e. $P(0)$ and $P(1)$ given by blue down-triangles and blue squares, respectively, with the numerical exact tMPS simulation, i.e. $P(0)$ and $P(1)$ represented by the blue solid line and blue dashed line, respectively. The quantum circuit simulation comprises chains of two-dimensional harmonic oscillators of length $l=12$. In coral red, we show its error as defined in the text.}
\label{fig:qttedopa}
\end{figure}

Lastly, we employed the Q-TEDOPA to simulate an environment with nonzero temperature. The environment is described by the temperature-dependent spectral density $J_{\beta}(\omega)$ defined in equation \eqref{jbeta}. We chose a temperature of $\text {T}=30$~K for the environment and an exponentially decaying Ohmic spectral density defined for positive and negative frequencies, as $J(\omega)=2\pi \alpha \omega e^{-\omega/\omega_{c}}$, where $\alpha=0.05$, $\omega_{c}=100cm^{-1}$ and the hard-cutoff frequencies are $\omega_{max}=-\omega_{min}=10\omega_{c}$ (more details in Appendix \ref{C}). We simulated the evolution of the whole system through classical simulation of the quantum circuits (without noise), which we compared to a numerically exact tMPS simulation (see Appendix \ref{D}). The simulations' results and error are shown in Figure \ref{fig:qttedopa}. The error of the quantum circuit simulation is $\approx 0.05$ at most for short time lengths. For longer evolution times, we expect higher error rates due to the truncated harmonic oscillator chain. To mitigate this effect, the number of oscillators in each chain may be increased or the chain Hamiltonian parameters may be optimized \cite{truncationTEDOPA}.
To conclude, we proposed, implemented and demonstrated with proof-of-concept simulations the quantum-computer-oriented versions of the TEDOPA and T-TEDOPA.
As the quantum hardware of NISQ devices improve, we expect the Q-TEDOPA to reach increasingly longer simulation times and become applicable to larger systems, making this quantum method an efficient alternative to simulate non-perturbative dynamics of generalized spin-boson models relatively to the HEOM or the conventional TEDOPA. We envisage that, in the future, it may be of great interest to apply the Q-TEDOPA to the work described in reference \cite{chin2013role}, so that the photosynthetic exciton transport can be simulated efficiently beyond dimer systems, i.e. for general light-harvesting complexes. Additional applications are Hamiltonian dynamics simulations of general quantum biological systems \cite{quantumeffectsbook,christensson} and condensed matter systems \cite{profmikhail,polariton,vibelectron} where perturbative approaches fail to provide full answers. In the future, it would be interesting to implement optimizations to reduce the harmonic oscillator chain length in the spirit of references \cite{truncationTEDOPA, nusseler2022fingerprint} and perhaps to find applications of the Q-TEDOPA to more general quantum computing problems besides quantum simulations.

\section{Acknowledgements}
The authors acknowledge helpful discussions and support by Jaemin Lim and Martin B. Plenio. J.D.G. acknowledges funding from the Portuguese Foundation for Science and Technology (FCT) through PhD Grant No. UI/BD/151173/2021 and support by the BMBF project PhoQuant (Grant No. 13N16110). Luís S. Barbosa was funded by National Funds through FCT within  the project IBEX (10.54499/PTDC/CCI-COM/4280/2021) and Mikhail I. Vasilevskiy acknowledges support in the framework of the Strategic Funding UIDB/04650/2020.

\bibliographystyle{quantum}

\onecolumn\newpage
\appendix

\section{Bosonic operator mapping} \label{sec:bosonic}
In this section, we show how to map bosonic creation and annihilation operators to Pauli operators using unary and binary qubit encondings for $d$-dimensional harmonic oscillators. We discuss the complexity of implementing the bosonic hopping term in the chain Hamiltonian \eqref{Hc2}, since the exponentiation of this term is the one that requires the highest number of CNOT gates on a quantum circuit. The mapping of the remaining Hamiltonian terms follows trivially from the former.

\subsection{Unary qubit encoding}
Each harmonic oscillator is represented by an array of $d$ qubits, where the leftmost qubit represents the energy level $m=0$ (ground state) and the rightmost qubit represents the $m=(d-1)$ energy level. The states $\ket{1}_{j}$ and $\ket{0}_{j}$ of the $j$th qubit represents the case where the harmonic oscillator is in the energy level $m=j$ and $m \neq j $, respectively. In this encoding, the phonon creation operator $b^{\dag}$ and phonon annihilation operator $b$ are given by,
\begin{equation}
    \begin{split}
    b^{\dag}=\sum_{j=0}^{d-1}\sqrt{j+1}\ket{0}\bra{1}_{j} \otimes \ket{1}\bra{0}_{j+1}, \\
    b=\sum_{j=0}^{d-1}\sqrt{j+1}\ket{1}\bra{0}_{j} \otimes \ket{0}\bra{1}_{j+1}.
\end{split}
\end{equation}

One can map the previous operators to Pauli operators as $\ket{0}\bra{1}=\frac{1}{2}(\hat{X}+i\hat{Y})$ and $\ket{1}\bra{0}=\frac{1}{2}(\hat{X}-i\hat{Y})$.

In reference \cite{bathoperators}, it has been shown that the unary qubit encoding of an operator with $O(d^2)$ non-zero entries, such as the bosonic hopping term in the Hamiltonian \eqref{Hc2}, requires, in the worst case, a number of CNOT gates that scales as $O(d^2)$. In practice, optimizations may be performed, e.g. removing equal adjacent gates or parallelizing the application of different Hamiltonian term's evolution operators across different qubits, effectively reducing the quantum circuit depth.

\subsection{Binary qubit encoding}
We represent each harmonic oscillator by an array of $d$ qubits, where the leftmost qubit corresponds to the Most Significant Bit (MSB) and the rightmost corresponds to the Least Significant Bit (LSB). Each state comprising the set of $log(d)$ qubits encodes an energy level $m$ of the harmonic oscillator, for instance $\ket{m=3}=\ket{0}\otimes \dots \otimes \ket{0} \otimes \ket{1}\otimes \ket{1}$. From this, it follows that,
\begin{equation}
\begin{split}
    b &= \sum_{m=0}^{d-1} \sqrt{m+1} \ket{m}\bra{m+1}, \\
    b^{\dag} &= \sum_{m=0}^{d-1} \sqrt{m+1}\ket{m+1}\bra{m}.
\end{split}
\end{equation}
The operators $\ket{m+1}\bra{m}$ and $\ket{m-1}\bra{m}$ must be converted to Pauli operators for each $m$.
One may efficiently obtain the Pauli strings of a term $b^{\dag}_{n}b_{n-1}+b^{\dag}_{n-1}b_{n}$ in binary qubit encoding for an arbitrary $d$ by constructing the sparse matrix of the operator and using a matrix-to-Pauli decomposition technique, such as the one available in the package \textit{Pennylane} \cite{pennylane}. Given the sparse representation of bosonic annihilation and creation operators, the decomposition of the bosonic hopping term in $H^{C}$, $b^{\dag}_{n}b_{n-1}+b^{\dag}_{n-1}b_{n}$,  takes, in a naive decomposition, about $O(d^4)$ classical operations. For $d=2$, the resulting decomposition is given as  $b^{\dag}_{n}b_{n-1}+b^{\dag}_{n-1}b_{n}=\frac{1}{2}\left[\hat{X}_{n}\hat{X}_{n-1}+\hat{Y}_{n}\hat{Y}_{n-1}\right]$.

As shown in reference \cite{bathoperators}, the bosonic hopping term in the Hamiltonian \eqref{Hc2} written in binary qubit encoding requires, in the worst case, a number of CNOT gates that scales as $O(d^2\log(d))$.

\subsection{Encoding scaling discussion}

Binary and unary encodings may be chosen based on the available quantum hardware to the user. As shown in Table \ref{tab:resources}, binary encoding allows one to exponentially reduce both the number of qubits and the required connections between adjacent qubits as a function of $d$, relatively to unary encoding. On the other hand, the circuit depth is increased by a factor of $log(d)$ (assuming full-qubit connectivity). In this work, binary qubit encoding was chosen due to the reduced qubit connectivity present in IBM quantum computers. Choosing the number of energy levels in each harmonic oscillator as $d=2$, only short-range qubit interactions (2-local) are required. Since the restricted qubit connectivity increases the number of required SWAP operators to implement the qubit interactions relatively to full-qubit connectivity platforms, it is relevant to notice that the implementation of 4-local interaction terms for unary qubit encoding would make necessary the use of SWAP operators. In the specific case of IBM platforms and $d=2$ oscillator encoding, unary encoding would increase the circuit depth relatively to the choice of binary encoding due to the nature of the qubits' interactions and the available quantum hardware.

\begin{table}
    \centering
    \begin{tabular}{||c|c|c||}
    \hline
     & Binary & Unary \\
    \hline
    \hline
      Qubits   & $\log(d)l$ & $dl$\\
      \hline
      Number of CNOT gates  & $O(lNd^2\log(d))$ & $O(lNd^2)$\\
      \hline
      Qubit connectivity & $2\log(d)$ & $2d$ \\
      \hline
    \end{tabular}
    \caption{Overview of the required quantum computational resources to implement Q-TEDOPA.}
    \label{tab:resources}
\end{table}

\subsection{Scaling of the commutator $\alpha_{comm}$ as a function of the chain's size} \label{order_comm}
We focus on the operator norm of the Trotter error decomposition $\alpha_{comm}=\sum_{\gamma_{1},\gamma_{2}=1}^{L}||[H_{\gamma_{2}},H_{\gamma_{1}}]||\leq O(l[d^2log(d)]^2)$ of the Hamiltonian $H^C_{E}$ shown in equation \eqref{Hc2}. The reason for this is that this term contains the highest number of Hamiltonian terms when decomposed to Pauli operators and subsequently applied in the quantum circuit (see Appendix subsections 1 and 2). We choose to do the calculation with the binary encoding, however we note that the calculation is straightforwardly generalized to unary encoding too.

Note that the number of terms in $H^C_{E}$ is $L = l d^2 log(d)$ for binary encoding. Consider that we first decompose the Hamiltonian in terms $H_{even}$ ($H_{odd}$) that are applied to even (odd) labelled oscillators, i.e. $H^C_{E} = H_{even} + H_{odd}$ as shown in Fig. \ref{fig:trotter_circuit}. Since the Hamiltonian comprises interactions only with nearest-neighbour qubits, we have $||[H_{even},H_{odd}]|| \leq O(l)$ since all terms that do not overlap over a qubit vanish. By noticing that each interaction term for a pair of nearest-neighbour oscillators in $H_{even}$ and $H_{odd}$ is decomposed into $d^2log(d)$ terms, one can include the dimension of the harmonic oscillator in the calculation, yielding the upper bound of the commutator as $||[H_{even},H_{odd}]|| \leq O(l[d^4log^2(d)])$.

\section{Resource estimates for previously reported classical simulations}\label{app: resource_est}
In the resource estimates, we neglect the implementation of the open system and system-environment interaction Hamiltonian terms since these require orders of magnitude less CNOT gates than the chain evolution Hamiltonian terms. We do not include any optimization to the quantum circuit and we neglect the counting of SWAP operators. For a summary of the resource estimate values, we refer the reader to Table \ref{tab:resource_est}.

The work in Ref. \cite{chin2013role} attempts to unveil the origins of coherence on the non-equilibrium process of the energy transport in photosynthesis using the T-TEDOPA. The authors try to simulate two molecules (encoded as qubits), each coupled to its own environment. Such model requires two chains of oscillators, each with $d=5$ energy levels per oscillator, chain size of $l=49$ and $N=188$ Trotter iterations. If the TEDOPA is implemented on a quantum device with binary (unary) encoding, then one needs $296$ ($786$) qubits and $\sim 3.5\times 10^{6}$ ($\sim 1.2\times 10^{6}$) CNOT gates for $d=8$ energy levels (i.e. $3$ qubits) per oscillator to run the evolution of the chain of oscillators. For the sake of comparison, let us consider that $d=4$ energy levels (i.e. $2$ qubits) per oscillator are sufficient to obtain accurate results. The resource estimates are then substantially reduced, roughly by half on the number of qubits and an order of magnitude on the number of CNOT gates, to $198$ ($394$) qubits and $\sim 5.9\times 10^{5}$ ($\sim 2.9\times 10^{5}$) CNOT gates using binary (unary) encoding. 

In Ref. \cite{dunnett2021influence}, the authors calculate the linear absorption spectra of the Methylene blue chromophore in aqueous solution using T-TEDOPA. The simulation requires one qubit (chromophore) coupled to three uncorrelated chains of harmonic oscillators with $d=20$, $l=150$ and $N=1000$. Two of the chains are coupled to the states of the chromophore, whereas the remaining one represents an environment in interaction with both states of the chromophore. This model, when implemented with Q-TEDOPA with binary (unary) encoding requires $2251$ ($14401$) qubits and $\sim 2.3 \times 10^{9}$ ($\sim 4.6 \times 10^{8}$) CNOT gates for $d=32$. On the other hand, choosing $d=16$, the resource estimates are once more substantially reduced, namely, $1801$ ($7201$) qubits and $4.6\times 10^7$ ($1.2\times 10^{7}$) CNOT gates for binary (unary) encoding.

\begin{table}[]
    \centering
    \begin{tabular}{||c|c|c|c||}
    \hline
    \multirow{2}{*}{Reference} & \multirow{2}{*}{Encoding} & \multirow{2}{*}{Qubits} & \multirow{2}{*}{CNOT gates}\\
    & & & \\
    \hline
    \hline
    \multirow{2}{*}{Ref. \cite{chin2013role}} & Binary & $296$ & $\sim 3.5\times 10^{6}$ \\
    \cline{2-4}
    & Unary & $786$ & $\sim 1.2\times 10^{6}$ \\
    \hline
    \multirow{2}{*}{Ref. \cite{dunnett2021influence}} & Binary & $2251$ & $\sim 2.3 \times 10^{9}$ \\
    \cline{2-4}
    & Unary & $14401$ & $\sim 4.6 \times 10^{8}$ \\
    \hline
    \end{tabular}
    \caption{Resource estimates for two types of open systems when simulated with Q-TEDOPA.}
    \label{tab:resource_est}
\end{table}

These estimates suggest that the choice of a small $d$ can yield a reduction of orders of magnitude of CNOT gates (i.e. circuit depth) and smaller numbers of qubits for the quantum simulation. Therefore, in general, low-medium temperature baths are better suit to the application of the Q-TEDOPA, because $d$ can be taken small.

\section{Quantum simulation numerical values} \label{sec:coeffs}
\subsection{System Hamiltonian parameters}

The excited state energies $\epsilon_{m}$ and electromagnetic coupling $g$ in the system Hamiltonian \eqref{HS} are characteristic of the first two molecules in the FMO complex of green sulphur bacteria (\textit{Cb. tepidum}) results reported in reference \cite{adolphs}. These are $\epsilon_{0}=12410$ $cm^{-1}$, $\epsilon_{1}=12530$ $cm^{-1}$ and $g=87.7$ $cm^{-1}$.

\subsection{Spectral density}

\begin{figure}[t]
\centering
\includegraphics[width=.7\textwidth]{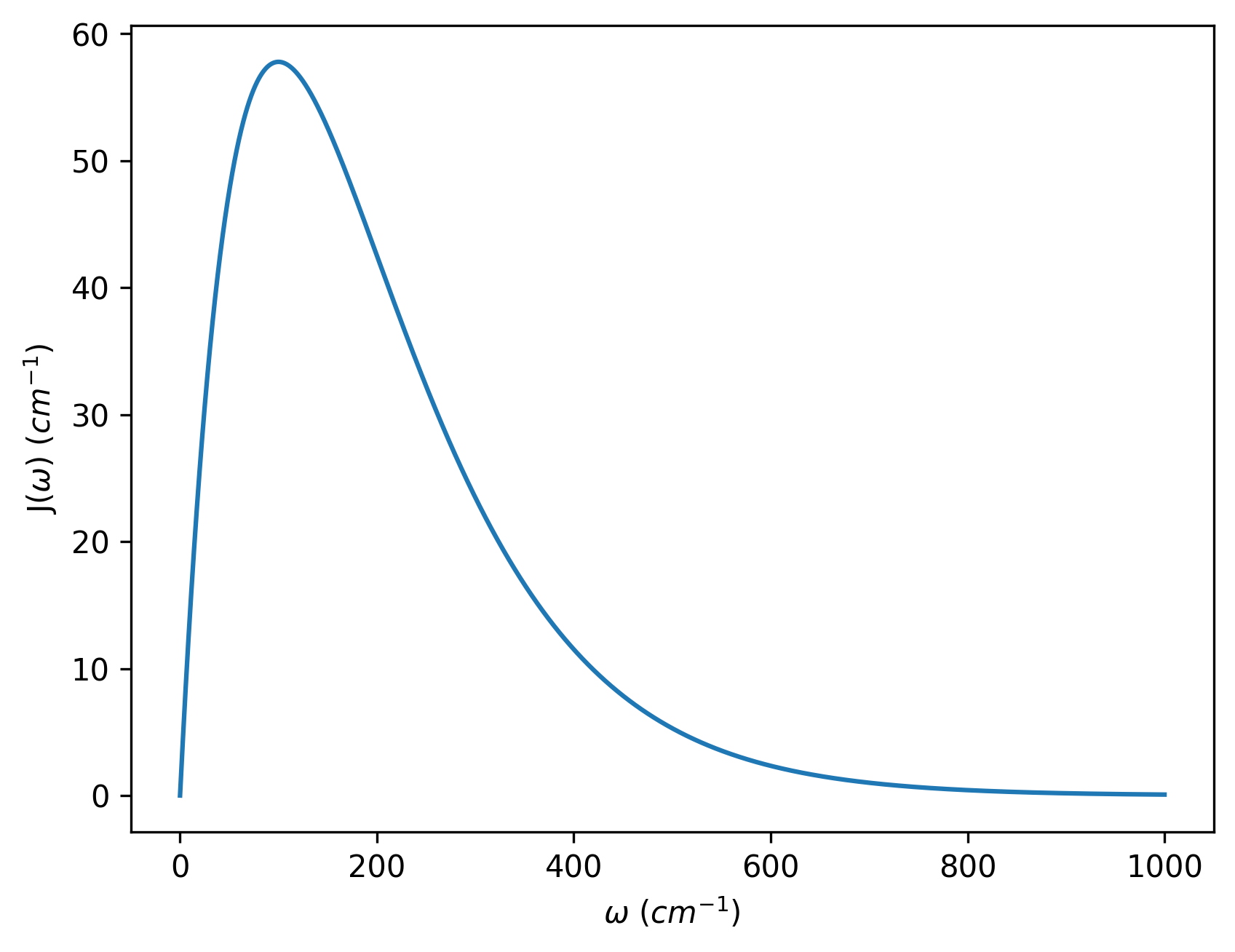}
\caption{Exponential decaying Ohmic  spectral density as per equation \eqref{SD}.}
\label{fig:SD}
\end{figure}

The spectral density in equation \eqref{SD} was chosen such that a moderate coupling strength between the open quantum system and the non-Markovian environment was used in the simulation. This spectral density allows us to highlight the main feature of TEDOPA and Q-TEDOPA, i.e. that it can simulate the evolution of an open quantum system in the regime where neither Förster or Redfield perturbative theories apply \cite{quantumeffectsbook}. We chose the hard frequency cutoff values of $\omega_{min}=0$ ($\text{T}=0K$) and $\omega_{max}=10\omega_{c}$ such that the missing environmental reorganization energy $\lambda_{miss} = \int_{\omega_{max}}^{\infty}d\omega J(\omega)/\omega$ is very small relatively to the total reorganization energy of the environment \cite{TEDOPA4}, $\lambda = \int_{0}^{\infty}d\omega J(\omega)/\omega$, i.e. $\lambda_{miss}/\lambda \approx 4.5 \times 10^{-5}$. For the spectral density $J_{\beta}(\omega)$ simulated in this work, one has $\lambda_{miss} =\int_{-\infty}^{\omega_{min}}d\omega J(\omega)/\omega+ \int_{\omega_{max}}^{\infty}d\omega J(\omega)/\omega$ and $\lambda = \int_{-\infty}^{\infty}d\omega J(\omega)/\omega$, from where we also obtained $\lambda_{miss}/\lambda \approx 4.5 \times 10^{-5}$. Therefore, the hard cutoff frequencies have a negligible impact on the accuracy of the simulations. We illustrate in Figure \ref{fig:SD} the spectral density \eqref{SD} used in this work.

\subsection{Values of the chain Hamiltonian coefficients}
Herein, we report the values of the chain Hamiltonian coefficients for the spectral density at $\text{T}=0$~K defined in equation \eqref{SD}.

The exciton-harmonic oscillator interaction coefficient is 
$t_{0}=70.69$ $cm^{-1}$.
\begin{center}
\begin{tabular}{| c | c | c |}
\hline
 $n$ & $w_{n}$ ($cm^{-1}$) & $t_{n+1,n}$ ($cm^{-1}$) \\ 
 \hline
 0 & 199.55 & 139.97 \\
 \hline
 1 & 385.14  & 222.93 \\
 \hline
 2 & 495.81  & 253.56 \\
 \hline
 3 & 514.13 & 253.63 \\
 \hline
 4 & 507.85 &  \\
 \hline
\end{tabular}
\end{center}

In the limit $n \to \infty$, one has $t_{n+1,n}=250$ $cm^{-1}$ and $w_{n}=500$ $cm^{-1}$ , implying that the the oscillators in the limit of a long $n$ encode the Markovian character of the bath, i.e. reflections of the excitations are harder to occur given the uniform distribution of oscillator couplings and energies, hence less excitations in the chain will travel back to interact again with the open system.

\section{Implementation of the quantum simulation}\label{C}
\subsection{Hardcore boson qubit encoding}
We restrict the dynamics of the open quantum system to the single-excitation Hilbert subspace where the relevant physics lies; hence excitons behave as hardcore bosons. The exciton creation and annihilation operators $c^{\dag}_{m}$ and $c_{m}$, respectively, act on a qubit as follows,
\begin{equation}
      \begin{split}
      c^{\dag}_{m}\ket{0}_{m}&=\ket{1}_{m}, \\
      c^{\dag}_{m}\ket{1}_{m}&=0, \\
      c_{m}\ket{0}_{m}&=0, \\
      c_{m}\ket{1}_{m}&=\ket{0}_{m}. \\
       \end{split}
 \end{equation}
 We map the excitonic operators to Pauli operators following a hardcore boson qubit encoding,
 \begin{equation}
 \begin{split}
    c^{\dag}_{m}=\ket{1}\bra{0}&=\frac{1}{2}(\hat{X}_{m}-i\hat{Y}_{m}), \\
    c_{m}=\ket{0}\bra{1}&=\frac{1}{2}(\hat{X}_{m}+i\hat{Y}_{m}), \\ 
    c^{\dag}_{m} c_{m} &= \frac{1}{2}(\hat{I}-\hat{Z}_{m}).
 \end{split}
 \end{equation}
 
 \subsection{Trotter-Suzuki product formula implementation}
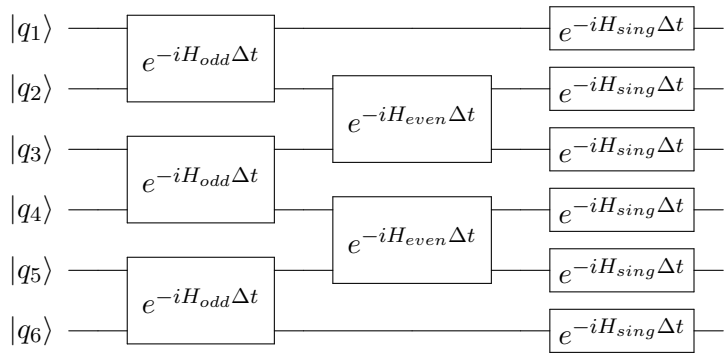
\begin{figure}[b]
\[\Qcircuit @C=1em @R=0.6em{ 
\lstick{\ket{q_{1}}} & \qw & \multigate{1}{e^{-iH_{odd}\Delta t}} & \qw & \qw & \qw & \gate{e^{-iH_{sing}\Delta t}} & \qw\\
\lstick{\ket{q_{2}}}& \qw & \ghost{e^{-iH_{odd}\Delta t}} & \qw & \multigate{1}{e^{-iH_{even}\Delta t}} & \qw & \gate{e^{-iH_{sing}\Delta t}} & \qw\\
\lstick{\ket{q_{3}}}& \qw & \multigate{1}{e^{-iH_{odd}\Delta t}} & \qw & \ghost{e^{-iH_{even}\Delta t}} & \qw& \gate{e^{-iH_{sing}\Delta t}} & \qw\\
\lstick{\ket{q_{4}}}& \qw &\ghost{e^{-iH_{odd}\Delta t}} & \qw & \multigate{1}{e^{-iH_{even}\Delta t}} & \qw & \gate{e^{-iH_{sing}\Delta t}} & \qw\\
\lstick{\ket{q_{5}}}& \qw  & \multigate{1}{e^{-iH_{odd}\Delta t}} & \qw & \ghost{e^{-iH_{even}\Delta t}} & \qw& \gate{e^{-iH_{sing}\Delta t}} & \qw\\
\lstick{\ket{q_{6}}}& \qw &\ghost{e^{-iH_{odd}\Delta t}} & \qw & \qw & \qw& \gate{e^{-iH_{sing}\Delta t}} & \qw\\
}\]
\centering
    \caption{Implementation of a Trotter iteration on a quantum circuit. The labelled qubits $q_{1}$, $q_{2}$, $q_{5}$ and $q_{6}$ are harmonic oscillators and the qubits $q_{3}$ and $q_{4}$ are the molecular qubits. The Trotter time step is $\Delta t = t/N$.}
    \label{fig:trotter_circuit}
\end{figure}

 The evolution operator is decomposed using a first-order Trotter-Suzuki product formula,
 \begin{equation}
     e^{-iHt}\approx \left[e^{-iH_{sing}t/N}e^{-iH_{even}t/N}e^{-iH_{odd}t/N}\right]^{N}.
 \end{equation}
 The simulated Hamiltonian is $H=H_{S}+H^{C}$ and $N$ is the number of Trotter iterations. We label the qubits in the quantum circuit as even and odd ones and apply the corresponding two-qubit Hamiltonian terms in $H$ as $H_{odd}$ and $H_{even}$, which include terms acting, orderly, on odd and even qubits or {\it vice versa}, respectively. The Hamiltonian terms $H_{sing}$ correspond to single-qubit operations, e.g. $c^{\dag}_{m}c_{m}$. Figure \ref{fig:trotter_circuit} illustrates the implementation of one iteration of the Trotter evolution for a $6$ qubit simulation with two-level harmonic oscillators. The generalization for simulations with $d$-level harmonic oscillators is performed by assigning the $even$ and $odd$ labels in the Hamiltonian terms to $even$ and $odd$ harmonic oscillators (in contrast to qubit assignment as explained above).
 
\subsection{State initialization, qubit measurement and quantum error mitigation}
 We initialize all harmonic oscillators in the vacuum state $\ket{0}$ and the molecular qubits in the state $\ket{10}$, i.e. first molecule in the excited state and the second molecule in the ground state. After evolving the system with $N$ Trotter iterations, we measure the probability of obtaining states $\ket{01}$ and $\ket{10}$ in the molecular qubits. We measure only these states in order to mitigate qubit amplitude damping errors. Excitons are conserved in the system, thus this error mitigation technique can be applied. One may admit excitation-preserving dynamics in the photosynthetic exciton transport if the dynamics we want to observe occurs on a timescale much smaller than exciton recombination processes. The transient coherent exciton transport dynamics, i.e. the physics that has the most interest to us (where quantum effects may play a fundamental role), occurs on a timescale less than $1$ $ps$ \cite{experiment1,experiment2}, which is about $\sim 10^{2}-10^{3}$ times smaller than typical exciton recombination processes \cite{FMOflow,quantumeffectsbook}.
 
\begin{figure}[h!]
\centering
\includegraphics[width=.9\textwidth]{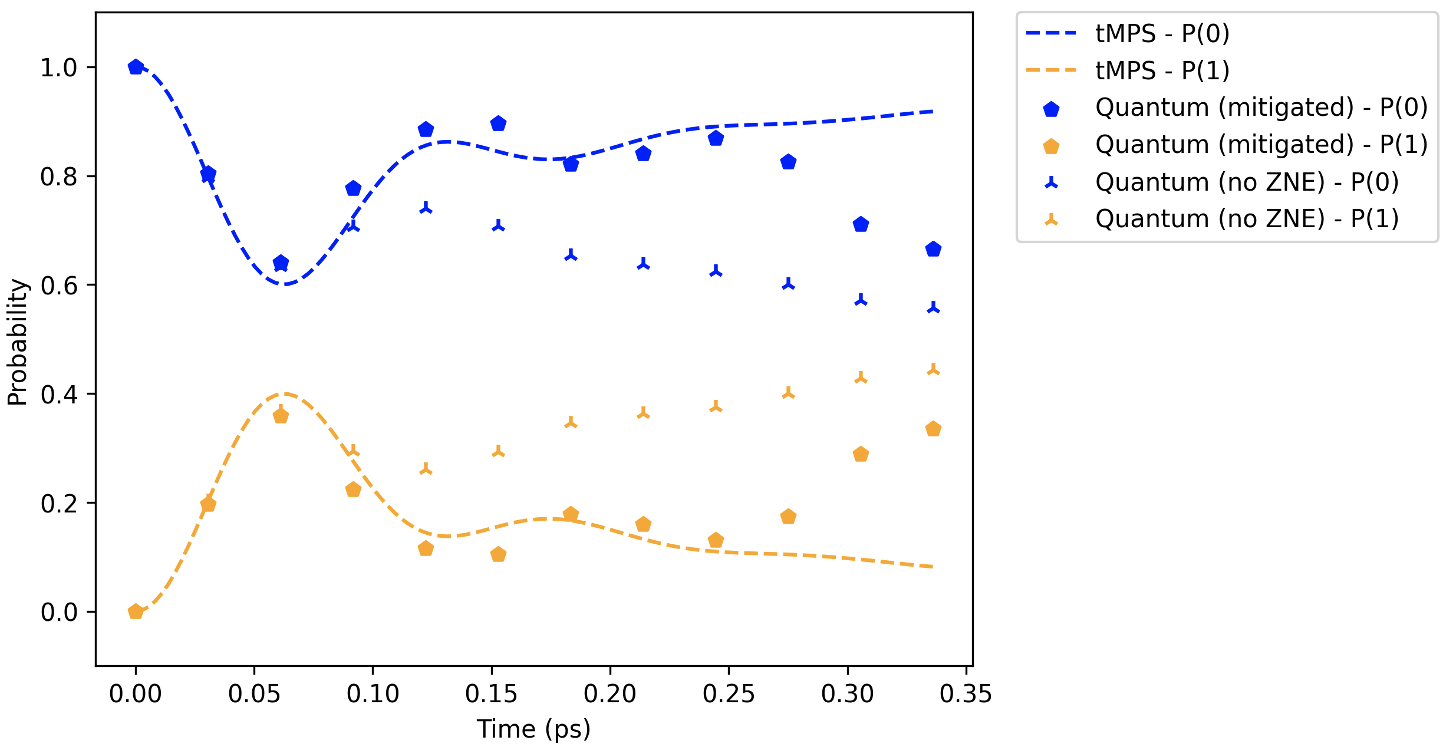}
\caption{Extended results (with 12 Trotter-Suzuki iterations) obtained with the quantum simulation in the \textit{ibmq toronto} with (without) the ZNE implemented, as represented by the pentagons (triangles). Numerically exact results obtained with tMPS (dashed lines) are also shown. The results obtained without ZNE contain all other error mitigation techniques referred in the text applied in the simulation.}
\label{fig:withoutzne}
\end{figure}

 We applied measurement error mitigation in \textit{Qiskit}, a technique that suppresses qubit readout errors.
 Two other effective error mitigation techniques were implemented in this work namely, the Randomized Compiling (RC) \cite{randcomp,randcomp2} and the Zero-Noise Extrapolation (ZNE) \cite{zne2,zne3,zne4}. The former suppresses coherent noise arising from under- and over- rotations of quantum gates by tailoring these errors into stochastic noise. Doing this, we are quadratically reducing the worst-case probability of obtaining an incorrect quantum gate implementation as a function of the average gate infidelity \cite{randcomp}. We applied RC to CNOT gates. The ZNE was also applied by folding the CNOT gates for $1$ (no folding) and $3$ times (via the Fixed Identity Insertion Method \cite{zne3}). For each CNOT folding factor, we implemented $RC$ by averaging the results of the $30$ randomized compiled quantum circuits, yielding a total of $2 \times 30$ executed quantum circuits for each simulation time $t$. Each circuit was measured $1.6\times 10^{4}$ times. ZNE allows the user to increase the average error rate in the quantum circuit and then extrapolate to the zero-noise result, hence mitigating incoherent noise in the quantum computer. We implemented exponential extrapolation \cite{zne2} using the software \textit{Mitiq} \cite{mitiq}. The impact of exponential ZNE in the quantum simulation results presented in Figure \ref{fig:results} for the $12$ simulated Trotter iterations is explicitly demonstrated in Figure \ref{fig:withoutzne}. We observed that exponential extrapolation is able to effectively suppress phase damping errors. However, it is unstable, whereas Richardson extrapolation is not very effective but stable for our particular quantum circuits \cite{zne2}. As shown recently in reference \cite{kim2023evidence}, the joint implementation of RC and ZNE increases the effectiveness of these quantum error mitigation techniques with respect to their individual application.
 
 \subsection{Noise in \emph{ibmq toronto}}
 In \textit{Qiskit}, one has access to the devices' qubit relaxation times and the duration the quantum circuit takes to be executed. We observed that the quantum digital simulation of $12$ Trotter iterations (instead of 10, shown in Fig.  \ref{fig:results}) with \textit{optimization level} $1$, pulses' scheduling method \textit{"as late as possible"}, and a system cycle time (\textit{ibmq toronto}) of $dt = 0.22$ $ns$ takes $\approx 47$ $\mu s$ to be executed. The average $T_{1}$ ($T_{2}$) relaxation time when we executed the quantum circuits was $\approx 82$ $\mu s$ ($\approx 103$ $\mu s$), therefore we expect non-negligible effects from qubit phase damping and, in particular, amplitude damping in the simulation results, as shown in Figure \ref{fig:withoutzne}. We remark that the limit of infinite noise is obtained when $P(0)=P(1)=0.5$ because we are measuring the molecular qubit states $\ket{01}$ and $\ket{10}$, both approximately equally affected by the amplitude damping noise.
 
  \begin{figure}[h!]
\centering
\includegraphics[width=.7\textwidth]{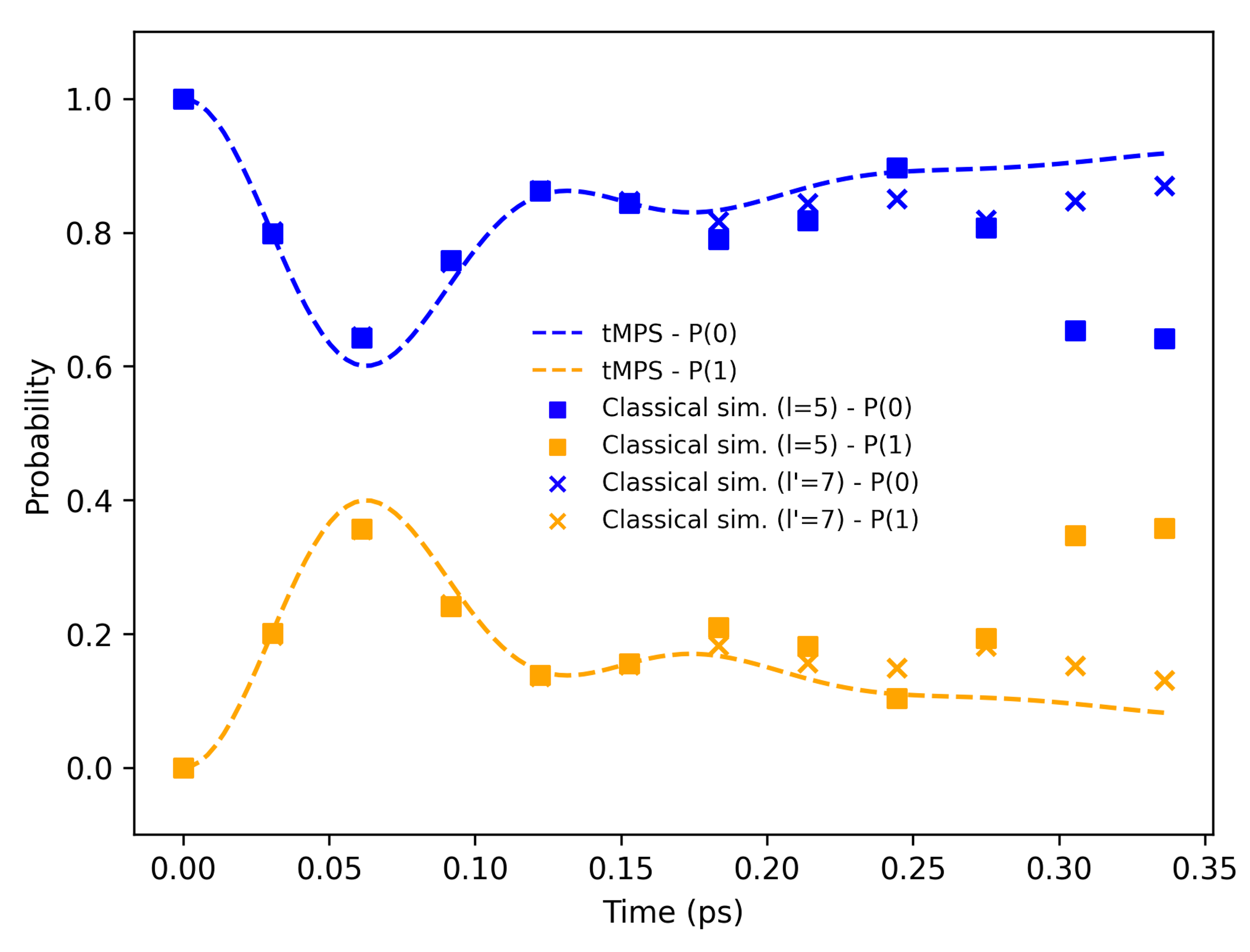}
\caption{Results obtained by the classical simulations of the quantum circuits (without noise). Results for a simulated chain length of $l=5$ ($l'=7$) are given by squares (crosses) and the numerical exact tMPS results are given by dashed lines.}
\label{fig:classical}
\end{figure}
 
 The decay of the last $3$ Trotter iterations in the results shown in Figure \ref{fig:withoutzne} (with ZNE implemented) is due to the truncated harmonic oscillator chain length. The abrupt truncation of the chain at $l=5$ induces a full reflection of the bosonic excitations that travel forward in the chain. Allowing enough time for the simulation, these artificial reflected excitations travel back through the chain again and interact with the system, producing unrealistic dynamics of the open system. This is demonstrated by the results in Figure \ref{fig:classical} obtained by classical simulations of the quantum circuits without noise. Given a simulation time length $t$, two simulations with different chain lengths, i.e. $l=5$ and $l'=7$, we observe that the open quantum system dynamics with the latter chain follow the numerical exact results up to the $12$th Trotter iteration, whereas in the former the effects of the reflection of the bosonic excitations at the edge of the truncated chain are clearly visible. Note that the unrealistic probability decay obtained in the results with $l=5$ is similar to the decay we observed in the quantum simulation results (with ZNE implemented) in Figure \ref{fig:withoutzne}, hence we conclude that the abrupt truncation of the chain introduces the artificial behaviour of the dynamics.
 
 \section{Matrix Product State simulation} 
 \label{D}
 The classically computed numerically exact results were obtained using the tMPS technique \cite{MPS1}. We decomposed the evolution operator using the second-order Trotter-Suzuki product formula and applied SVD compression to each resulting Matrix Product Operator, allowing for a truncation error $\varepsilon>0$, upper bounded by $\varepsilon<10^{-4}$. The Matrix Product States were compressed after each Trotter iteration with SVD, where we allowed a truncation error upper bounded by $\varepsilon<10^{-8}$. The TEDOPA (T-TEDOPA) simulation of the environment at $\text {T}=0$ ($\text {T}=30$~K) consisted of chains of harmonic oscillators with dimension $d=8$ ($d=8$), chains' length $l=15$ ($l=35$) and $120$ ($160$) Trotter iterations.

\end{document}